\documentstyle[12pt]{article}
\newcommand{\beq}{\begin{equation}}
\newcommand{\eeq}{\end{equation}}
\newcommand{\beqa}{\begin{eqnarray}}
\newcommand{\eeqa}{\end{eqnarray}}
\newcommand{\beqan}{\begin{eqnarray*}}
\newcommand{\eeqan}{\end{eqnarray*}}
\newcommand{\no}{\nonumber}

\newcommand{\ul}{\underline}
\newcommand{\ol}{\overline}
\newcommand{\ra}{\rightarrow}
\newcommand{\lra}{\longrightarrow}

\newcommand{\ben}{\begin{enumerate}}
\newcommand{\een}{\end{enumerate}}
\newcommand{\bfl}{\begin{flushleft}}
\newcommand{\efl}{\end{flushleft}}
\newcommand{\ba}{\begin{array}}
\newcommand{\ea}{\end{array}}
\newcommand{\btab}{\begin{tabular}}
\newcommand{\etab}{\end{tabular}}
\newcommand{\bit}{\begin{itemize}}
\newcommand{\eit}{\end{itemize}}

\newcommand{\vs}{\vspace}
\newcommand{\hs}{\hspace}
\newcommand{\prepr}[1] {\begin{flushright} {\bf #1} \end{flushright} \vskip
1.5cm}
\newcommand{\titul}[1] {\begin{center}{\large \bf #1 } \end{center}\vskip 1.cm}

\newcommand{\autor}[1] {\begin{center} {\bf \lineskip .3cm #1  }
                        \end{center} }

\newcommand{\lugar}[1] {\begin{center}  {\large \it #1   } \end{center}}
\newcommand{\abstr}[1] {{\begin{center} \vskip .5cm {\bf Abstract
                        \vspace{0pt}} \end{center}}\begin{quote} #1
                        \end{quote}}
\newcounter{muni}

\begin{document}
\vspace{.4cm}
\hbadness=10000
\pagenumbering{arabic}
\begin{titlepage}
\prepr{Preprint hep-ph/940XXXX \\PAR/LPTHE/94-30  \\21 July 1994 }
\titul{ Implications of factorization for the determination of \\
 hadronic form factors in $D_s^+ \ra \phi $ transition.}
\autor{ M. Gourdin\footnote{\rm Postal address: LPTHE, Tour 16, $1^{er}$ Etage,
Universit\'e Pierre {\it \&} Marie Curie and Universit\'e Denis Diderot, 4
Place Jussieu, F-75252 Paris CEDEX 05, France.},
\hs{3mm} A. N. Kamal$^{1,}\footnote{Permanent address : Department of Physics,
University of Alberta, Edmonton, Alberta T6G2JI, Canada }$ }
\autor{ Y. Y. Keum$^1$ and X. Y. Pham$^1$ }

\lugar{Laboratoire de Physique Th\'eorique et Hautes Energies,\footnote{\em
Unit\'e associ\'ee au CNRS URA 280}
 Paris, France }

\begin{center}
{\small  E-mail : gourdin@lpthe.jussieu.fr, keum@lpthe.jussieu.fr,
pham@lpthe.jussieu.fr, \\ kamal@lpthe.jussieu.fr and
kamal@bose.phys.ualberta.ca}
\end{center}

\vs{-12cm}
\thispagestyle{empty}
\vs{120mm}
\noindent
\abstr{
Using factorization we determine the allowed domains of the ratios of form
factors, $x = A_2(0)/A_1(0)$ and $y = V(0)/A_1(0)$, from the experimentally
measured ratio $R_h \equiv \Gamma(D_s^+ \ra \phi \rho^+)/\Gamma(D_s^+ \ra \phi
\pi^+)$
assuming three different scenarios for the $q^2$-dependence of the form
factors.
We find that the allowed domains overlap with those obtained by using the
experimentally measured ratio $R_{s\ell} = \Gamma(D^+_s \ra \phi \ell^+
\nu_{\ell})/\Gamma(D^+_s \ra \phi \pi^+)$ provided that the phenomenological
parameter $a_1$ is $1.23$.
Such a comparison presents a genuine test of factorization.
We calculate the longitudinal polarization fraction, $\Gamma_L/\Gamma \equiv
\Gamma(D_s^+ \ra \phi_L \rho^+_L)/\Gamma(D_s^+ \ra \phi \rho^+)$, in the three
scenarios for the $q^2$-dependence of the form factors and emphasize the
importance of measuring $\Gamma_L/\Gamma $.
Finally we discuss the $q^2$-distribution of the semileptonic decay and find
that  it is rather insensitive to the scenarios  for the $q^2$-dependence of
the form factors, and unless very accurate data can be obtained
it is unlikely to discriminate between the different scenarios.
Useful information on the value of $x$ might be obtained by the magnitude of
the $q^2$-distribution near $q^2 = 0$.
However the most precise information on $x$ and $y$ would come from the
knowledge of the longitudinal and left-right transverse polarizations of the
final vector mesons in hadronic and/or semileptonic decays.
}
%
\end{titlepage}

\newpage
Since the Cabibbo-Kobayashi-Maskawa(CKM)-favoured decays of the strange-charmed
meson $D_s^+$ involve, at the quark level, $s\ol{s}$ pair, the vector meson
$\phi$ emerges naturally as one of the decay products.
Among its decay modes are the three most important ones, $D_s^+ \ra \phi
\pi^+$, $D_s^+ \ra \phi \rho^+$ and $D_s^+ \ra \phi \ell^+ \nu_{\ell} (\ell =
e, \mu)$ measured by several groups \cite{With}, and the topic of this paper.

The analysis of the experimental data for these modes involves three axial form
factors, $A_0$, $A_1$ and $A_2$, and a vector form factor $V$, in the notation
of Bauer, Stech and Wirbel (BSW) \cite{BSW}.
In spite of this fact which renders the analysis of experimental data involving
the $\phi$ meson more complicated than the corresponding one with pseudoscalar
mesons such as $\eta$ or $\eta'$, it turns out that the polarization of the
$\phi$ meson may provide
crucial information that can be used to test the factorization assumption which
is one of the basic ingredients in the computation of hadronic decay rates.

The fact that the $\pi$ meson is light allows us  to ignore the variation of
the hadronic form factor $A_0^{D_s\phi}$ between $q^2 = m_{\pi}^2$, where it
enters the description of $D_s^+ \ra \phi\pi^+$, and $q^2 = 0$.
This then allows us to eliminate $A_0^{D_s\phi}(0)$ in favour of
$A_1^{D_s\phi}(0)$ and $A_2^{D_s\phi}(0)$ through \cite{BSW},
\beq
2 m_{\phi} A_0^{D_s\phi}(0) = (m_{D_s} + m_{\phi}) A_1^{D_s\phi}(0) - (m_{D_s}
- m_{\phi}) A_2^{D_s\phi}(0) \label{eq1}
\eeq

Since the other decays, $D_s^+ \ra \phi \rho^+$ and $D_s^+ \ra \phi \ell^+
\nu_{\ell}$, involve the form factors $A_1^{D_s\phi}$, $A_2^{D_s\phi}$ and
$V^{D_s\phi}$, it proves convenient to factor out $A_1^{D_s\phi}(0)$ which then
cancels in the ratio of the rates, and define the following ratios of the form
factors at $q^2 = 0$,
\beq
x = \frac{A_2^{D_s\phi}(0)}{A_1^{D_s\phi}(0)} \, \hs{20mm} y =
\frac{V^{D_s\phi}(0)}{A_1^{D_s\phi}(0)} \ \label{eq2}
\eeq

In our analysis we do not adopt any of the several models of hadronic form
factors for $D_s^+ \ra \phi$ transition available in the literature to compute
decay rates as such a procedure would not tell whether the factorization
assumption or the model-form factors were being tested in the comparison
between theory and experiment.
Instead, we simply consider different scenarios for the $q^2$-variation of the
form factors ( and do not make any assumptions about their absolute
normalizations ) and show that the measured ratio of the two hadronic rates,
\beq
R_h \equiv \Gamma(D_s^+ \ra \phi \rho^+)/\Gamma(D_s^+ \ra \phi \pi^+)
\label{eq3a}
\eeq
constrains the parameters $x$ and $y$ to an allowed domain of the $x$, $y$
plane ( limited by $x$ and $y$ non-negative a suggested by all phenomenological
models ).
Such a domain can then be compared with the values of $x$ and $y$ already
measured in the semileptonic modes \cite{{With},{E-653},{E-687}}.
We remind the reader that semileptonic data are analyzed with monopole
$q^2$-dependence for all form factors which is just one of the scenarios we
discuss.
A procedure such as ours that eschews reliance on a particular set of
model-form factors produces not only a more bias-free test of factorization but
also one that is independent of the BSW phenomenological parameter $a_1$
\cite{BSW}.

Further, since the $\phi$ meson is longitudinally polarized in $D_s^+ \ra \phi
\pi^+$, the ratio $\Gamma(D_s^+ \ra \phi \pi^+)/\Gamma(D_s^+ \ra \phi \rho^+)$
behaves physically like the longitudinal polarization fraction
$\Gamma_L/\Gamma$ in $D_s^+ \ra \phi \rho^+$, where
\beq
\frac{\Gamma_{L}}{\Gamma} \ \equiv \frac{\Gamma(D_s^+ \ra \phi_L
\rho_L^+)}{\Gamma(D_s^+ \ra \phi \rho^+)} \ \label{eq3}
\eeq

Therefore, there must be a link between these two quantities $R_h$ and
$\Gamma_L/\Gamma$  which we explicitly demonstrate.
Indeed, an upper and a lower limit for $\Gamma_L/\Gamma$ can be calculated for
the allowed $x$, $y$ domain.
A direct measurement of this quantity would provide a further test of
factorization.

We now recall, for completeness, the various expressions for decay rates used
in this paper though they can be found in published literature
\cite{{Pham1},{Gilman},{Pham2}}.
We start with the semileptonic decay $D_s^+ \ra \phi \ell^+ \nu_{\ell}$ in the
zero lepton mass limit.
Let us denote by $q^2$ the (effective mass)$^2$ of the $\ell \nu_{\ell}$-system
and introduce the following dimensionless quantities,
\beq
r \equiv \frac{m_{\phi}}{m_{D_s}} \, \hs{20mm} t^2 \equiv \frac{q^2}{m^2_{D_s}}
\ \label{eq4}
\eeq
with $ 0 \leq t^2 \leq (1 - r)^2$.

In the $D_s^+$ rest system, the $\phi$ meson momentum $K(t^2)$ is given by,
\beq
K(t^2) = \frac{m_{D_s}}{2}\ k(t^2) \label{eq5}
\eeq
where $k(t^2) = [(1 + r^2 - t^2)^2 - 4 r^2]^{1/2}$.

The  $t^2$-distribution for semileptonic decay is the sum of the contributions
from the three polarization states,
\beq
\frac{d\Gamma}{dt^2} \ (D_s^+ \ra \phi \ell^+ \nu_{\ell}) =
\frac{G_F^2 m^5_{D_s}}{192 \pi^3} \ \hs{2mm}|V_{cs}|^2 \hs{2mm} k(t^2) \hs{2mm}
\{H_{LL}(t^2) + H_{++}(t^2) + H_{--}(t^2)\} \label{eq6}
\eeq
where
\beqa
H_{LL}(t^2) & = & [\frac{1 + r}{2 r} \ ]^2 \hs{2mm} |(1 - r^2 - t^2)
A_1^{D_s\phi}(t^2) - \frac{k^2(t^2)}{(1 + r)^2} \ A_2^{D_s\phi}(t^2)|^2 \no, \\
H_{\pm\pm}(t^2) & = & (1 + r)^2 \hs{2mm} t^2 \hs{2mm} |A_1^{D_s\phi}(t^2) \mp
\frac{k(t^2)}{(1 + r)^2} \ V^{D_s\phi}(t^2)|^2 \label{eq7}
\eeqa

For the hadronic modes where only the spectator diagram contributes in the
factorization approximation, we obtain

\beq
\Gamma(D_s^+ \ra \phi \pi^+) = \frac{G_F^2 m^5_{D_s}}{32 \pi} \ |V_{cs}|^2
|V_{ud}|^2 \hs{2mm} a_1^2 \hs{2mm} ( \frac{f_{\pi^+}}{m_{D_s}} \ )^2 \hs{2mm}
k(0) \hs{2mm} H_{LL}(0) \label{eq8}
\eeq
where for t = 0,

\beq
k(0) \hs{2mm} H_{LL}(0) = (1 - r^2)^3 \hs{2mm} [ \frac{1 + r}{2 r} \ ]^2
\hs{2mm} \{1 - \frac{1 - r}{1 + r} \ x \}^2 \hs{2mm} |A_1^{D_s\phi}(0)|^2
\label{eq9}
\eeq

For the $\phi \rho^+$ mode, first treating the $\rho$ meson as a zero-width
resonance, we get
\beq
\Gamma(D_s^+ \ra \phi \rho^+) = \frac{G_F^2 m^5_{D_s}}{32 \pi} \ |V_{cs}|^2
|V_{ud}|^2 \hs{2mm} a_1^2 \hs{2mm} ( \frac{f_{\rho^+}}{m_{D_s}} \ )^2 \hs{2mm}
k(t_{\rho}^2) \hs{2mm}
\{H_{LL}(t_{\rho}^2) + H_{++}(t_{\rho}^2) + H_{--}(t_{\rho}^2) \}
 \label{eq10}
\eeq
where $t_{\rho} = m_{\rho}/m_{D_s} $.
A comparison of (9) and (11) provides us with the precise connection between
$\Gamma(D_s^+ \ra \phi \pi^+)$ and $\Gamma(D_s^+ \ra \phi_L \rho_L^+)$,
the rate for the $\rho$ meson in the longitudinally polarized state.

To take the finite $\rho$-width into account one has to smear the rate given in
(11) over $t^2$ with a Breit-Wigner measure \cite{Pham2},

\beq
k(t_{\rho}^2) \hs{2mm} H_{\lambda\lambda}(t_{\rho}^2) \lra \frac{1}{\pi} \
\int_{\frac{4 m_{\pi}^2}{m^2_{D_s}} \ }^{(1 - r)^2}
dt^2
\frac{ \gamma_{\rho} t_{\rho} }{(t^2 - t_{\rho}^2)^2 + \gamma_{\rho}^2
t_{\rho}^2} \ k(t^2) H_{\lambda\lambda}(t^2) \label{eq11}
\eeq
where $\gamma_{\rho} = \Gamma_{\rho}/m_{D_s} $, $\Gamma_{\rho}$ being the
$\rho$-width.

The analysis of the experimental data requires certain assumptions for the
$q^2$-dependence of the hadronic form factors.
In our analysis we assume the vector form factor, $V(q^2)$, to have a monopole
$q^2$-dependence with a pole mass $m_{1^-} =2.11 \hs{2mm} GeV$ \cite{BSW}.
As for the axial form factors, $A_1(q^2)$ and $A_2(q^2)$, we investigate the
following three scenarios :

\vs{3mm}
Scenario I : Both $A_1(q^2)$ and $A_2(q^2)$ assumed to have monopole
$q^2$-dependence with pole mass $m_{1^+} =2.53 \hs{2mm}GeV$ as in BSW
\cite{BSW}.

Scenario II : Both $A_1(q^2)$ and $A_2(q^2)$ assumed to be independent of $q^2$
in the region of interest.

Scenario III : $A_1(q^2)$ assumed to be decreasing with $q^2$,
\beq
A_1^{D_s\phi}(q^2) = A_1^{D_s\phi}(0)[1 - \frac{q^2}{(3.5 \hs{2mm}GeV)^2} \ ]
\label{eq12}
\eeq
while $A_2(q^2)$ is assumed to have a monopole dependence with a pole mass of
$3.5 \hs{2mm}GeV$.
This latter value was chosen just for illustration of a gently decreasing and
increasing $A_1(q^2)$ and $A_2(q^2)$ respectively.
Qualitatively, Scenarios II and III were inspired by the QCD sum rule
calculations of Ref. \cite{Ball} for $D \ra K^*$ transition form factors.

Using the measured \cite{Cleo1} hadronic ratio,
\beq
R_h = \frac{\Gamma(D_s^+ \ra \phi \rho^+)}{\Gamma(D_s^+ \ra \phi \pi^+)}
=1.86 \pm 0.26 \hs{2mm} ^{ + 0.29}_{ - 0.40} \label{eq13}
\eeq
we calculated the allowed domains in the $x$,$y$ plane for each of the above
scenarios.
For $\Gamma(D_s^+ \ra \phi \pi^+)$ we used expressions (9) and (10), and for
$\Gamma(D_s^+ \ra \phi \rho^+)$ the $t^2$-smeared form of (11) in accordance
with the prescription of Eq.(12).

In figs. 1, 2(a) and 2(b) we have plotted the boundary curves for scenarios I,
II and III respectively. All positive values of $x$ and $y$ are allowed up to
the boundary curve.
Experiments \cite{{E-653},{E-687}} on semileptonic decays, $D_s^+ \ra \phi
\ell^+ \nu_{\ell}$, have determined $x$ and $y$ using the assumptions of
scenario I. We have plotted their points in Fig. 1 too.
We note that the point from E-653 \cite{E-653} is excluded by the constraint
set by $R_h$ while the point from E-687 \cite{E-687} has a small overlap with
the domain determined from $R_h$.
In our analysis the two errors in (14) were added in quadrature and we used the
leptonic decay constants $f_{\pi^+} = 131.7 \hs{2mm} MeV$ and $f_{\rho^+} = 212
\hs{2mm} MeV$ in Eqs. (9) and (11).

Technically the curves for fixed $R_h$ are hyperbolae in the $x$, $y$ plane
centered on the $x$ axis whose principal axes are the $x$ axis and an axis
parallel to the $y$ one.
Of course only the part of the curve located on the left branch of the
hyperbola has a physical meaning; the second branch requires too large a value
of $x$, $x \geq 5.0$.

The general equation for these hyperbolae is,

\beq
\frac{(x - x_c)^2}{a^2} \ - \frac{y^2}{b^2} \ = 1. \label{eq14}
\eeq
We have listed the numerical values of $x_c$, $a$ and $b$ for the three models
in Table 1 where we have also tabulated the upper limits of $x$ and $y$,
$x_{max}$ for $y$ = 0 and $y_{max}$ for $x$ = 0, and, of course, $x_{max}$ =
$x_c$ - $a$.

\vs{5mm}
\begin{center}
\begin{tabular}{|c||c|c|c|c|c|} \hline
SCENARIO & $x_c $  & $a$ &  $b$ & $x_{max}$ & $y_{max}$ \\
\hline\hline
I & 2.81 & 2.37 & 6.14 & 0.44 & 3.92 \\
\hline
II & 2.87 & 2.13 & 5.57 & 0.74 & 5.01 \\
\hline
III & 2.89 & 2.00 & 5.20 & 0.89 & 5.45 \\
\hline
\end{tabular} \\

\vs{5mm}
$\ul{\rm Table \hs{2mm} 1.}$ \vs{2mm}

{\small Parameters of the hyperbolae and $x_{max}$ and $y_{max}$ of the allowed
$x$, $y$ domain.}

\end{center}

The full curves in Figs. 1, 2(a) and 2(b) use the smeared rate for
$\Gamma(D^+_s \ra \phi \rho^+)$ in Eq.(12) while the dotted curves use the
$\rho^+$ zero-width approximation in Eq.(11).
We remark that for scenario I the allowed $x$, $y$ domain for the $\rho^+$
zero-width approximation is empty. We emphasise that the corrections due to the
non-zero $\rho$ meson width play an important role in all scenarios.

Having established the range of $x$ and $y$ allowed by $R_h$, we can calculate
using (11) and (12), the maximum and the minimum values of the longitudinal
polarization fraction. We find the following limits;
\beqa
{\rm scenario \hs{3mm} I } \hs{6mm}: \hs{20mm} & & 0.43 \leq
\frac{\Gamma_L}{\Gamma} \ \leq 0.55 \no \\
{\rm scenario \hs{3mm} II } \hs{4mm}: \hs{20mm} & & 0.36 \leq
\frac{\Gamma_L}{\Gamma} \ \leq 0.55 \label{eq15}\\
{\rm scenario \hs{3mm} III } \hs{2mm}: \hs{20mm} & & 0.33 \leq
\frac{\Gamma_L}{\Gamma} \ \leq 0.55 \no
\eeqa

The upper limit occurs for $x = y = 0$ and is insensitive to the scenario
adopted, while the lower limit depending on the maximum value of $y$, which
occurs at $x$ = 0, is rather scenario-dependent.
In passing, we note that for a zero-width $\rho$-meson the maximum value of
$\Gamma_L/\Gamma$ is 0.51.

It is evident that an experimental measurement of the vector meson polarization
in the decay $D^+_s \ra \phi \rho^+$ could provide important information on the
parameters $x$ and $y$.
Further, a large experimental value of $\Gamma_L/\Gamma$ exceeding
unambiguously the upper bound given in (16) will be a clear indication of the
failure of the factorization approximation.
Equally, a measurement of $\Gamma_L/\Gamma$ below 0.33 would rule out all the
three scenarios  assuming, of course, the validity of factorization
approximation.
In short, $\Gamma_L/\Gamma$ could be used to discriminate between different
scenarios.

In Figs. 1, 2(a) and 2(b) we have also plotted constant $\Gamma_L/\Gamma$
curves (dashed-dotted) for three values, 0.3, 0.4 and 0.5 in order to exhibit
the variation of $\Gamma_L/\Gamma$ in the $x$, $y$ plane.
As previously, these curves are also hyperbolae of the type given by Eq.(15).

We now turn to a discussion of the semileptonic decays $D^+_s \ra \phi \ell^+
\nu_{\ell}$.
Experimentally, the following ratio has been measured
\cite{{With},{E-687},{Cleo2},{Argus},{Cleo3}}
\beq
R_{s\ell} \equiv \frac{\Gamma(D^+_s \ra \phi \ell^+ \nu_{\ell})}{\Gamma(D^+_s
\ra \phi \pi^+ )} \hs{2mm} = \hs{2mm} 0.54 \pm 0.10 \label{eq16}
\eeq

Using (7) and (8) we can calculate $\Gamma(D^+_s \ra \phi \ell^+ \nu_{\ell})$
for each of our three scenarios for the $q^2$-dependence of the form factors.
Then using (9) and (10) we can construct a theoretically $a_1$-independent
ratio,
\beq
a_1^2 \hs{2mm} \frac{\Gamma(D^+_s \ra \phi \ell^+ \nu_{\ell})}{\Gamma(D^+_s \ra
\phi \pi^+ )} = a_1^2 \hs{2mm} R_{s\ell} \label{eq17}
\eeq
If $a_1$ were a well-determined parameter we could use (18) with $R_{s\ell} $
given by (17) to determine the allowed $x$, $y$ domain from the ratio
$R_{s\ell}$.
It turns out that the limits of the allowed domains are rather sensitive to the
value of $a_1$ used.
If we treat $a_1$ as a free parameter we find that for $a_1 \approx 1.23$ the
allowed domains obtained from $R_h$ and $R_{s\ell}$ almost overlap.
This is quite remarkable as the expected value of $a_1$ is $1.26$
\cite{Neubert}.
This we interpret as an evidence in favour of the factorization hypothesis in
the hadronic decays.

\vs{5mm}
We now address the question : What can we learn from the $q^2$-distribution,
$d\Gamma(D^+_s \ra \phi \ell^+ \nu_{\ell})/dq^2$, of the semileptonic decays ?

For $0^- \ra 0^- \ell \nu_{\ell}$ decay the answer is simple - we determine the
$q^2$-dependence of the form factor $F_1(q^2)$ as there is only one  form
factor involved. In the present case of $0^- \ra 1^- \ell \nu$ decay the answer
is not so clear as there are four terms in the $q^2$-distribution associated
with $A_1^2(q^2), A_1(q^2) \cdot A_2(q^2), A_2^2(q^2)$ and $V^2(q^2)$ and the
only way to disentangle these four contributions, and then to measure the
$q^2$-dependence of the various form factors, is to determine simultaneously
the longitudinal and left-right transverse polarizations of the final vector
meson. We are far from achieving this in the case of
$D^+_s \ra \phi \ell^+ \nu_{\ell}$ decay in the near future.

Unfortunately,  experimental measurement of the true $q^2$-distribution of the
semileptonic decays has not yet been possible because the momentum of the
$D^+_s$ decaying in flight is not known \cite{With1} which, in turn, implies
that the $q^2$ for an event is not accurately known.
However, two experiments, E-653 \cite{E-653} and E-687 \cite{E-687} have
determined the parameters $x$ and $y$, assuming scenario I for the
$q^2$-dependence of the form factors, from the semileptonic decay $D^+_s \ra
\phi \ell^+ \nu_{\ell}$.
In the following we exploit this knowledge in absence of information on the
$q^2$-distribution $d\Gamma/dq^2$.

Let us introduce a unit normalized dimensionless quantity $X(t^2)$,
\beq
X(t^2) = \frac{1}{\Gamma} \
\frac{d}{dt^2} \ \Gamma(D^+_s \ra \phi \ell^+ \nu_{\ell}) \label{eq18}
\eeq
which depends on $x$ and $y$, and on the type of $q^2$-variation of the
hadronic form factors.
We first construct an "experimental" $X(t^2)$, denoted by $\ol{X}(t^2)$ in the
following, using the experimental values of $x$ and $y$ obtained from a triple
angular fit of the data \cite{{E-653},{E-687}} in the scenario I.
We then compute the same quantity, $X(t^2)$, for the values of $x$ and $y$
inside the allowed domains determined by us from $R_h$ and shown in Figs. 1,
2(a) and 2(b).
Finally we compare the $X(t^2)$ so constructed with $\ol{X}(t^2)$.

{}From Fig. 1 we note that the E-653 point lies well outside the allowed $x$,
$y$ domain while the E-687 point has a small overlap with this domain.
Consequently we use only the latter data point in constructing $\ol{X}(t^2)$
and we adopt \cite{E-687},
\beq
x = 1.1 \pm 0.81, \hs{20mm} y = 1.8 \pm 0.92 \label{eq19}
\eeq
 using  scenario I, as was also the case in \cite{E-687}.

The resulting $\ol{X}(t^2)$ is shown in Fig. 3(a) in the form of nine curves
corresponding to the central and one standard deviation extreme values of $x$
and $y$.
These nine curves break up naturally into three groups of three curves each,
with each group being characterized by the value of $x$ ; within each group the
splitting due to different values of $y$ is a much finer effect.
Qualitatively at $t^2 = 0$, the group with the smallest $x$ has the largest
intercept with the vertical axis and within each group the curve with the
smallest $y$ has the largest intercept.

Our calculated $X(t^2)$ for scenarios I, II and III are shown in Figs. 3(b),
4(a) and 4(b) respectively in the form of three curves corresponding to the
three extreme values of $x$ and $y$ for the allowed domains of Figs. 1, 2(a)
and 2(b), $(x,y) = (0,0), (0,y_{max}), (x_{max},0)$.
All other choices of $x$ and $y$ in the allowed domain yield $X(t^2)$ which
lies within the envelope of the three curves shown in Figs. 3(b), 4(a) and
4(b).
The $t^2$-distribution in these figures are then our prediction for $X(t^2)$
for $x$ and $y$ obtained from the hadronic ratio $R_h$.
We observe that they depend strongly on $x$ and very little on the $q^2$
scenarios.
Therefore, we conclude that $X(t^2)$ cannot distinguish between different $q^2$
scenarios unless very precise measurements of the $q^2$-distribution can be
made
which probably will not be the case in the near future.
However, the determination of $X(t^2)$ near $t^2 = 0$ appears to be very useful
for the determination of the parameter $x$.

We now compare our curves $X(t^2)$ with the "experimental" ones, $\ol{X}(t^2)$,
of Fig. 3(a). For scenario I, we find some overlap between the curves of Fig.
3(a) and 3(b) provided that $x$ is small.
In particular the group of curves in Fig. 3(a) with $x = 0.29 $ overlap well
with the curve in Fig. 3(b) which uses $x = x_{max}$ and $y = 0 $.
The reason is obvious from Fig. 1 which shows that the points, $x = 0.29$ and y
between 1 and 2 lie inside the allowed domain of $x$ and $y$ obtained from
$R_h$.
We remind the reader that $X(t^2)$ depends only weakly on $y$.

Such an overlap appears a little more problematic in scenarios II and III
especially for large values of $t^2$ but cannot be excluded.
For this reason we suggest to experimentalists to fit their triple angular
distributions with different scenarios for the $q^2$-dependence of the hadronic
form factors for extraction  of $x$ and $y$.
It would be interesting to see if the corresponding extracted values of $x$ and
$y$ for scenarios II and III would be consistent with the allowed domains of
$x$ and $y$ for these scenarios extracted from $R_h$ and shown respectively in
Figs.  2(a) and 2(b).

In conclusion, assuming factorization and using $R_h$ (see(3) and (14) for
definition), appropriately smeared over the $\rho$-width, we found the allowed
domains in the $x$, $y$ plane using three different scenarios for the
$q^2$-dependence of the form factors.
These allowed domains plotted in Figs 1, 2(a) and 2(b) show sensitivity to the
choice of the scenario.
There is an overlap, though small, with the E-687 \cite{E-687} values of $x$
and $y$ from semileptonic decay $\Gamma(D^+_s \ra \phi \ell^+ \nu_{\ell})$ and
the $x$, $y$ allowed domain calculated by us from $R_h$ in the relevant
scenario - scenario I ( See Fig. 1).
This would suggest that the hadronic data from \cite{Cleo1} and the
semileptonic data from \cite{E-687} are consistent with the factorization
hypothesis.
E-653 \cite{E-653} determination of $x$ and $y$ from semileptonic decay, on the
other hand, does not support factorization assumption.

We have also determined the allowed domains of $x$ and $y$ from $R_{s\ell}$
(see (17) for definition) in the three scenarios.
Such a determination is $a_1$-dependent.
We found that the two domains determined from $R_h$ and $R_{s\ell}$ overlap
almost completely for $ a_1 \approx 1.23 $ in remarkable agreement with the
theoretically computed value of $a_1 = 1.26$ \cite{Neubert}.
This fact supports the validity of the factorization assumption.

We have plotted the contours of constant $\Gamma_L/\Gamma$ ( see (4) for
definition)
in the $x$, $y$ plane and calculated the maximum and the minimum values of
$\Gamma_L/\Gamma$ for the three scenarios of $q^2$-dependence of the form
factors considered by us.
It is remarkable that the maximum value of $\Gamma_L/\Gamma$ = 0.55 is
independent of the scenario for the $q^2$-dependence of the form fators, while
the minimum value is scenario-dependent. Of course, a measurement of the
longitudinal and left-right transverse polarizations will determine both $x$
and $y$ in a chosen scenario for the $q^2$-dependence of the form factors.

A measurement of $\Gamma_L/\Gamma$ could be important in testing factorization.
For example, should the measured value of $\Gamma_L/\Gamma$ exceed 0.55
unambiguously, it would signal a break down of the factorization hypothesis.
A value below 0.33 would rule out all of our three scenarios for the
$q^2$-dependence of the form factors if factorization were to hold.

Finally, we suggest that experimentalists analyze their semileptonic data using
for $q^2$-dependence of the form factor forms other than monopoles to extract
the value of $x$ and $y$.

%
%
%
%

\vspace{1cm}
\hspace{1cm} \Large{} {\bf Acknowledgements}    \vspace{0.5cm}

\normalsize{
A. N. K wishes to thank the Laboratoire de Physique Th\'eorique et Hautes
Energies
in Paris for their hospitality and the Natural Sciences and Engineering
Research Council of Canada
for a research grant which partly supported this research.

Y. Y. K would like to thank the Commissariat \`a l'Energie Atomique of France
for the award of a fellowship
and especially G. Cohen-Tannoudji for encouragements.
}

\newpage
%

\newpage
\section*{Figure captions}
\normalsize
\vspace{0.5cm}

\begin{enumerate}

\item
{\bf Fig. 1} : \hs{3mm}
The shaded region bounded by the solid line is the allowed domain of $x$, $y$
from $R_h$ within one standard deviation.
The dashed-dot lines correspond to fixed values of $\Gamma_L/\Gamma $ as
indicated. All these curves are calculated with the scenario I for the $q^2$
dependence of form factors. \\

\item
{\bf Fig. 2(a) and 2(b)} : \hs{3mm}
Same as Fig. 1, for scenarios II and III respectively.
The dotted lines replace the solid lines in the zero width approximation of the
$\rho$ meson. \\

\item
{\bf Fig. 3} :
\begin{enumerate}
\item
 "Experimental" $\ol{X}(t^2)$ as explained in the text. \\
\item
 Calculated  $X(t^2)$ in the scenario I. \\
\end{enumerate}

\item
{\bf Fig. 4} :
\begin{enumerate}
\item
 Calculated  $X(t^2)$ in the scenario II. \\
\item
 Calculated  $X(t^2)$ in the scenario III.
\end{enumerate}

\end{enumerate}


\begin{thebibliography}{99}
%

\bibitem{With}
M. S. ~Witherell, Invited talk at the International Symposium on Lepton and
Photon Interactions at High Energies, Cornell, Ithaca, N. Y.,
 Report No. UCSB-HEP-93-06, 1993 (to be published).
%
\bibitem{BSW} M. ~Wirbel, B. ~Stech and M. ~Bauer, {\em Z. Phys.} {\bf C29} 637
(1985);
M. ~Bauer, B. ~Stech and M. ~Wirbel, {\em Z. Phys.} {\bf C34} 103 (1987).
%
\bibitem{E-653}
K. ~Kodama et al.(E-653), {\em Phys. Lett.} {\bf B309}, 483(1993).
%
\bibitem{E-687}
P. L. ~Frabetti et al.(E-687), {\em Phys. Lett.} {\bf B328}, 187(1994).
%
\bibitem{Pham1}
R. ~Nabavi, X. Y. ~Pham and W. N. ~Cottingham, {\em J. Phys.} {\bf G3},
1485(1977); X. Y. ~Pham and J. -M. Richard, {\em Nucl. Phys.} {\bf B138},
453(1978).
%
\bibitem{Gilman}
J. G. ~K\"orner and G. A. ~Schuler, {\em Z. Phys.} {\bf C46}, 93(1990);
F. J. Gilman and R. L. Singleton, {\em  Phys. Rev.} {\bf D41}, 142(1990).
%
\bibitem{Pham2}
X. Y. ~Pham and X. C. ~Vu, {\em Phys. Rev.} {\bf D46}, 261(1992).
%
\bibitem{Ball}
P. ~Ball, V. M. ~Braun and H. G. ~Dosch, {\em Phys. Rev.}  {\bf D44},
3567(1991).
%
\bibitem{Cleo1}
P. ~Avery et al.(CLEO), {\em Phys. Rev. Lett.} {\bf 68}, 1279(1992).
%
\bibitem{Cleo2}
J. ~Alexander et al.(CLEO), {\em Phys. Rev. Lett.} {\bf 65}, 1531(1990).
%
\bibitem{Argus}
H. ~Albrecht et al.(ARGUS), {\em Phys.  Lett.} {\bf B255}, 634(1991).
%
\bibitem{Cleo3}
F. ~Butter et al.(CLEO), {\em Phys. Lett.} {\bf B324}, 255(1994).
%
\bibitem{Neubert}
M.~Neubert, V.~Rieckert, B.~Stech and Q.P.~Xu, in  $\ul{Heavy \hs{2mm}
Flavours}$,
 Eds. A. J. Buras and M. Lindner, ( World Scientific, Singapore, 1992 ).
%
\bibitem{With1}
M. S. ~Witherell, Private Communication.

\end{thebibliography}
\end{document}